\documentstyle[preprint,aps,epsfig]{revtex}

\begin{document}

%\draft

\preprint{
\hfill$\vcenter{\hbox{\bf IFT-P.081/98}
}$ }

\title{Signal and Backgrounds for Leptoquarks at the LHC II: \\
Vector Leptoquarks}

\author{A.\ Belyaev$^{1,2}$\thanks{Email: belyaev@ift.unesp.br},
O.\ J.\ P.\ \'Eboli$^1$\thanks{Email: eboli@ift.unesp.br}, 
R.\ Z.\ Funchal$^3$\thanks{Email: zukanov@charme.if.usp.br}, 
and T.\ L.\ Lungov$^1$\thanks{Email: lungov@ift.unesp.br} }

\address{\em $^1$ Instituto de F\' {\i}sica Te\'orica -- UNESP \\
            R. Pamplona 145, 01405-900 S\~ao Paulo, Brazil \\
            $^2$ Skobeltsyn Institute of Nuclear Physics, 
              Moscow State University \\
              119899 Moscow, Russian Federation \\
          $^3$ Instituto de F\'{\i}sica, Universidade de S\~ao Paulo \\
    C.P.\ 66.318, 05315-970 S\~ao Paulo, Brazil}

\maketitle

\vskip -18pt

\begin{abstract}
  
\vskip -18pt

  We perform a detailed analyses of the CERN Large Hadron Collider
  (LHC) capability to discover first generation vector leptoquarks
  through their pair production.  We study the leptoquark signals and
  backgrounds that give rise to final states containing a pair
  $e^+e^-$ and jets.  Our results show that the LHC will be able to
  discover vector leptoquarks with masses up to 1.8--2.3 TeV depending
  on their couplings to fermions and gluons.

\end{abstract}


\newpage

%%%%%%%%%%%%%%%%%%%%%%%%%%%%%%%%%%%%%%%%%%%%%%%%%%%%%%%%%%%%%%%%%%%%%%
\section{Introduction}

In the standard model (SM) the cancellation of the chiral anomalies
takes place only when we consider the contributions of leptons and
quarks, indicating a deeper relation between them.  Therefore, it is
rather natural to consider extensions of the SM that treat quarks and
leptons in the same footing and consequently introduce new bosons,
called leptoquarks, that mediate quark-lepton transitions.  The class
of theories exhibiting these particles includes composite models
\cite{comp,af}, grand unified theories \cite{gut}, technicolor models
\cite{tec}, and superstring-inspired models \cite{e6}. Since leptoquarks 
couple to a lepton and a quark, they are color triplets under
$SU(3)_C$, carry simultaneously lepton and baryon number, have
fractional electric charge, and can be of scalar or vector nature.

From the experimental point of view, leptoquarks possess the striking
signature of a peak in the invariant mass of a charged lepton with a
jet, which make their search much simpler without the need of
intricate analyses of several final state topologies.  Certainly, the
experimental observation of leptoquarks is an undeniable signal of
physics beyond the SM, so there have been a large number of direct
searches for them in $e^+e^-$ \cite{LEP}, $e^\pm p$ \cite{HERA}, and
$p \bar{p}$ \cite{PP} colliders.  Up to now all of these searches led
to negative results, which bound the mass of vector leptoquarks to be
larger than 245--340 (230--325) GeV, depending on the leptoquark
coupling to gluons, for branching ratio into $e^\pm$-jet equal to 1
(0.5) \cite{donote}.

The direct search for leptoquarks with masses above a few hundred GeV
can be carried out only in the next generation of $pp$ \cite{fut:pp},
$ep$ \cite{buch,fut:ep}, $e^+e^-$ \cite{fut:ee}, $e^-e^-$
\cite{fut:elel}, $e\gamma$ \cite{fut:eg}, and $\gamma\gamma$
\cite{fut:gg} colliders. In this work, we extend our previous analyses 
of the LHC potentiality to discover scalar leptoquarks to vector ones
\cite{nos}. We study the pair production of first generation 
leptoquarks that lead to a final state topology containing two jets
plus a pair $e^+e^-$. We analyze this signal for vector leptoquarks
and use the results for the SM backgrounds obtained in Ref.\
\cite{nos}, where careful studies of all possible top production, 
QCD and electroweak
backgrounds for this topology were performed using the event generator
PYTHIA \cite{pyt}. We restrict ourselves to first generation
leptoquarks that couple to pairs $e^\pm u$ and $e^\pm d$ with the
leptoquark interactions described by the most general effective
Lagrangian invariant under $SU(3)_C \otimes SU(2)_L \otimes U(1)_Y$
\cite{buch}.

In this work, we study the pair production of vector leptoquarks via
quark-quark and gluon-gluon fusions, {\em i.e.}
\begin{eqnarray}
q + \bar{q}~ &&\rightarrow \Phi_{\text{lq}} + \bar{\Phi}_{\text{lq}} \; ,
\label{eq:qq}
\\
g + g~ &&\rightarrow \Phi_{\text{lq}} + \bar{\Phi}_{\text{lq}} \; ,
\label{eq:gg}
\end{eqnarray}
where we denote the vector leptoquarks by $\Phi_{\text{lq}}$. These processes
give rise to $e^+e^-$ pairs with large transverse momenta accompanied by jets.
Using the cuts devised in Ref.\ \cite{nos} to reduce the backgrounds and
enhance the signals, we show that the LHC will be able to discover first
generation vector leptoquarks with masses smaller than 1.5--2.3 TeV, depending
on their couplings and on the integrated luminosity (10 or 100 fb$^{-1}$).

Here, we perform our analyses using a specially created event generator for
vector leptoquarks.  Moreover we consider the most general coupling of vector
leptoquarks to gluons, exhibiting our results for two distinct scenarios. In
particular we analyze the most conservative case where the leptoquark
couplings to gluons is such that the pair production cross section is 
minimal.~\cite{bbk}.
While we were preparing this paper, a similar study of the production of
vector leptoquarks appeared \cite{dion}, which uses a different event
generator, distinct cuts and a less general leptoquark coupling to gluons,
which contains only the chromomagnetic anomalous coupling to gluons.

Low energy experiments give rise to strong constraints on leptoquarks, unless
their interactions are carefully chosen \cite{shanker,fcnc}. In order to evade
the bounds from proton decay, leptoquarks are required not to couple to
diquarks. To avoid the appearance of leptoquark induced FCNC, leptoquarks are
assumed to couple only to a single quark family and only one lepton
generation. Nevertheless, there still exist low-energy limits on leptoquarks.
Helicity suppressed meson decays restrict the couplings of leptoquarks to
fermions to be chiral \cite{shanker}. Moreover, residual FCNC \cite{leurer},
atomic parity violation \cite{apv}, effects of leptoquarks on the $Z$ physics
through radiative corrections \cite{jkm} and meson decay
\cite{leurer,apv,davi} constrain the first generation leptoquarks to be
heavier than $0.5$--$1.5$ TeV when the coupling constants to fermions are
equal to the electromagnetic coupling $e$. Therefore, our results indicate
that the LHC can not only confirm these indirect limits but also expand them
considerably.

The outline of this paper is as follows. In Sec.\ \ref{l:eff} we
introduce the $SU(3)_C \otimes SU(2)_L \otimes U(1)_Y$ invariant
effective Lagrangians that we analyzed. In Sec.\ \ref{mc} we describe
in detail how we have performed the signal Monte Carlo simulation.
Sec.\ \ref{bckg} contains a brief summary of the backgrounds and
kinematical cuts needed to suppress them.  Our results and conclusions
are shown in Sec.\ \ref{resu}.

%%%%%%%%%%%%%%%%%%%%%%%%%%%%%%%%%%%%%%%%%%%%%%%%%%%%%%%%%%%%%%%%%%%%%%
\section{Models for vector leptoquark interactions}
\label{l:eff}

In this work we assume that leptoquarks decay exclusively into the known
quarks and leptons. In order to avoid the low energy constraints, leptoquarks
must interact with a single generation of quarks and leptons with chiral
couplings. Furthermore, we also assume that their interactions are $SU(3)_C
\otimes SU(2)_L \otimes U(1)_Y$ gauge invariant above the electroweak symmetry
breaking scale $v$. The most general effective Lagrangian satisfying these
requirements and baryon number (B), lepton number (L), electric charge, and
color conservations is \cite{buch}~:
\begin{eqnarray}
{\cal L}^{\text{f}}_{\text{eff}}~  &  & =~ {\cal L}_{F=2} ~+~ {\cal L}_{F=0} 
+ \text{h.c.} \; , 
\label{e:int}
\\
{\cal L}_{F=2}~  &  & =~ g_{\text{2L}}~ (V^{\mu}_{2L})^T~ \bar{d}^c_R~ 
\gamma_\mu~ i \tau_2~ \ell_L +
g_{\text{2R}}~ \bar{q}^c_L~ \gamma_\mu~i \tau_2~ e_R ~ V^{\mu}_{2R} 
+ \tilde{g}_{\text{2L}}~ (\tilde{V}^{\mu}_{2L})^T 
\bar{u}^c_R~ \gamma_\mu~ i \tau_2~ \ell_L  
\; ,
\label{lag:fer}\\
{\cal L}_{F=0}~  &  & =~ h_{\text{1L}}~ \bar{q}_L~ \gamma_\mu~ \ell_L~ 
V^{\mu}_{1L} 
+ h_{\text{1R}}~ \bar{d}_R~  \gamma_{\mu}~ e_R~ V^{\mu}_{1R} 
+ \tilde{h}_{\text{1R}}~ \bar{u}_R~ \gamma_{\mu}~ e_R~ \tilde{V}^{\mu}_{1R}
+ h_{\text{3L}}~ \bar{q}_L~ \vec{\tau}~ \gamma_{\mu}~ \ell_L \cdot 
\vec{V}^{\mu}_{3L}
\; ,
\label{eff} 
\end{eqnarray}
where $F=3B+L$, $q$ ($\ell$) stands for the left-handed quark (lepton)
doublet, and $u_R$, $d_R$, and $e_R$ are the singlet components of the
fermions. We denote the charge conjugated fermion fields by
$\psi^c=C\bar\psi^T$ and we omitted in Eqs.\ (\ref{lag:fer}) and
(\ref{eff}) the flavor indices of the leptoquark couplings to
fermions. The leptoquarks $V^{\mu}_{1R(L)}$ and $\tilde{V}^{\mu}_{1R}$
are singlets under $SU(2)_L$, while $V^{\mu}_{2R(L)}$ and
$\tilde{V}^{\mu}_{2L}$ are doublets, and $V^{\mu}_{3L}$ is a triplet.

From the above interactions we can see that for first generation leptoquarks,
the main decay modes of leptoquarks are those into pairs $e^\pm q$ and $\nu_e
q^\prime$. In this work we do not consider their decays into neutrinos,
however, we take into account properly the branching ratio into charged
leptons.  In Table \ref{t:cor} we exhibit the leptoquarks that can be studied
using the final state $e^{\pm}$ plus a jet, as well as their decay products
and branching ratios.  Only the leptoquarks $V^{2}_{2L}$,
$\tilde{V}^{2}_{2L}$, and $V_3^{+}$ decay exclusively into a jet and a
neutrino, and are not constrained by our analyses; see Eqs.\ (\ref{lag:fer})
and (\ref{eff}).

Leptoquarks are color triplets, therefore, it is natural to assume that they
interact with gluons. However, the $SU(2)_C$ gauge invariance is not enough to
determine the interactions between gluons and vector leptoquarks since it is
possible to introduce two anomalous couplings $\kappa_g$ and $\lambda_g$ which
are related to the anomalous magnetic and electric quadrupole moments
respectively. We assume here that these quantities are independent in order to
work with the most general scenario.  The effective Lagrangian describing the
interaction of vector leptoquarks ($\Phi$) with gluons is given by \cite{bbk}
\begin{equation}
\label{eqLAV}
{\cal L}_V^g
=   -\frac{1}{2} V^{i \dagger}_{\mu \nu}
V^{\mu \nu}_i + M_\Phi^2 \Phi_{\mu}^{i \dagger} \Phi^{\mu}_i - 
ig_s \left [ (1 - \kappa_g)
\Phi_{\mu}^{i \dagger}
t^a_{ij}
\Phi_{\nu}^j
{\cal G}^{\mu \nu}_a
+ \frac{\lambda_g}{M_\Phi^2} V^{i\dagger}_{\sigma \mu}
t^a_{ij}
V_{\nu}^{j \mu} {\cal G}^{\nu \sigma}_a \right ] ,
\end{equation}
where there is an implicit sum over all vector leptoquarks, $g_s$
denotes the strong coupling constant, $t^a$ are the $SU(3)_C$
generators, $M_\Phi$ is the leptoquark mass, and $\kappa_g$ and
$\lambda_g$ are the anomalous couplings, assumed to be real.  The field
strength tensors of the gluon and vector leptoquark fields are
respectively
\begin{eqnarray}
{\cal G}_{\mu \nu}^a  &=& \partial_{\mu} {\cal A}_{\nu}^a
 - \partial_{\nu}
{\cal A}_{\mu}^a + g_s f^{abc} {\cal A}_{\mu b} {\cal A}_{\nu c},
 \nonumber\\
V_{\mu \nu}^{i}
 &=& D_{\mu}^{ik}
 \Phi_{\nu k} - D_{\nu}^{ik} \Phi_{\mu k},
\end{eqnarray}
with the covariant derivative given by
\begin{equation}
D_{\mu}^{ij} = \partial_{\mu} \delta^{ij} - i g_s
t_a^{ij}
 {\cal A}^a_{\mu} \; ,
\end{equation}
where ${\cal A}$ stands for the gluon field.

At present there are no direct bounds on the anomalous parameters $\kappa_g$
and $\lambda_g$. Here we analyze two scenarios: in the first, called minimal
cross section couplings, we minimize the production cross section as a
function of these parameters for a given vector leptoquark mass.  In the
second case, which we name Yang--Mills couplings, we consider that the vector
leptoquarks are gauge bosons of an extended gauge group which corresponds to
$\kappa_g=\lambda_g=0$.

%%%%%%%%%%%%%%%%%%%%%%%%%%%%%%%%%%%%%%%%%%%%%%%%%%%%%%%%%%%%%%%%%%%%%%
\section{Signal Simulation and Rates}
\label{mc}

Although the processes for the production of scalar leptoquarks are
incorporated in PYTHIA, the vector leptoquark production is absent.  In order
to study the pair production of vector leptoquarks via the processes
(\ref{eq:qq}) and (\ref{eq:gg}) we have created a Monte Carlo generator for
these reactions, adding a new external user processes to the
PYTHIA~5.7/JETSET~7.4 package \cite{pyt}. We have included in our simulation
two cases of anomalous vector leptoquark couplings to gluons, as well as their
decays into fermions.

In our analyses, we assume that the pair production of leptoquarks is due
entirely to strong interactions, {\i.e.}, we neglect the contributions from
$t$-channel lepton exchange via the leptoquark couplings to fermions
\cite{bbk}.  This hypothesis is reasonable since the fermionic couplings $g$
and $h$ are bounded to be rather small by the low energy experiments for
leptoquarks masses of the order of TeV's.

The analytical expressions for the scattering amplitudes were taken from the
{\bf \sf LQPAIR} package \cite{lqpair}, which was created using the {\sf
  CompHEP} package \cite{comphep}. The
integration over the phase space was done using {\bf \sf BASES} \cite{bases}
while we used {\bf \sf SPRING} for the simulation \cite{bases}. An interface
between these programs and PYTHIA was specially written.

In our calculations we employed the parton distribution functions CTEQ3L
\cite{cteq2l}, where the scale $Q^2$ was taken to be the leptoquark mass
squared.  Furthermore, the effects of final state radiation, hadronization and
string jet fragmentation (by means of JETSET~7.4) have also been taken into
account.

\begin{figure}
\begin{center}
\mbox{\epsfig{file=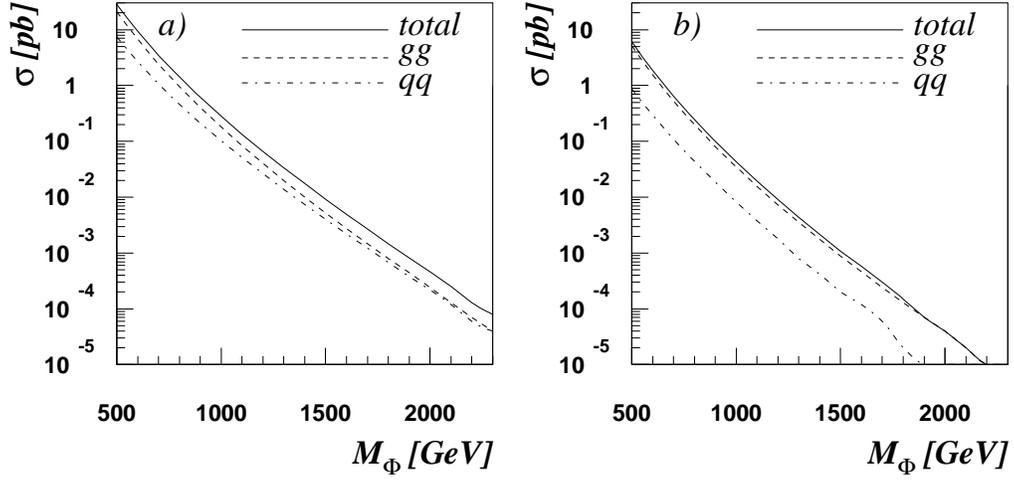,width=0.9\textwidth}}
\end{center}
\caption{Production cross sections of vector leptoquarks pairs
  at the LHC for (a) Yang--Mills coupling and (b) minimum coupling (cross
  section).}
\label{f:pair1}
\end{figure}

The cross sections for the production of vector leptoquark pairs are presented
in Fig.\ \ref{f:pair1} for Yang--Mills and minimal couplings.  The numerical
values of the total cross sections are shown in Table \ref{t:pair1} along with
the values of couplings $\kappa_g$ and $\lambda_g$ that lead to the minimum
total cross section. As we can see from this figure, the gluon-gluon fusion
mechanism (dashed line) dominates the production of leptoquark pairs for the
leptoquark masses relevant for this work at the LHC center-of-mass energy.
Moreover, quark--quark fusion is less important in the minimal coupling
scenario.

Pairs of leptoquarks decaying into $e^\pm$ and a $u$ or $d$ quark produce a
pair $e^+ e^-$ and two jets as signature.  In our analyses we kept track of
the $e^\pm$ (jet) carrying the largest transverse momentum, that we denoted by
$e_1$ ($j_1$), and the $e^\pm$ (jet) with the second largest $p_T$, that we
called $e_2$ ($j_2$). Furthermore, we mimicked the experimental resolution of
the hadronic calorimeter by smearing the final state quark energies according
to
\begin{eqnarray*}
\left.\frac{\delta E}{E}\right|_{had} &=& \frac{0.5}{\sqrt{E}} \; .
\end{eqnarray*}
The reconstruction of jets was done using the subroutine LUCELL of PYTHIA.
The minimum $E_T$ threshold for a cell to be considered as a jet initiator has
been chosen 2 GeV, while we assumed the minimum summed $E_T$ for a collection
of cells to be accepted as a jet to be 7 GeV inside a cone $\Delta R =\sqrt{
  \Delta \eta^2 + \Delta \phi^2} =0.7$.  The calorimeter was divided on $(50
\times 30)$ cells in $\eta \times \phi$ with these variables in the range
$(-5<\eta<5) \times (0.<\phi<2\pi)$.

%%%%%%%%%%%%%%%%%%%%%%%%%%%%%%%%%%%%%%%%%%%%%%%%%%%%%%%%%%%%%%%%%%%%%%
\section{Background Processes and Kinematical Cuts}
\label{bckg}

Within the scope of the SM, there are many sources of backgrounds leading to
jets accompanied by a $e^+e^-$ pair, which we classify into three classes
\cite{nos}: QCD processes, electroweak interactions, and top quark production.
The reactions included in the QCD class depend exclusively on the strong
interaction and the main source of hard $e^\pm$ in this case is the
semileptonic decay of hadrons possessing quarks $c$ or $b$.  The electroweak
processes contains the Drell--Yan production of quark pairs and the single and
pair productions of electroweak gauge bosons.  Due to the large gluon-gluon
luminosity at the LHC, the production of top quark pairs is important by
itself due to its large cross section. These backgrounds have been fully
analyzed by us in Ref.\ \cite{nos} and we direct the reader to this reference
for further information.

In order to enhance the signal and reduce the SM backgrounds we have devised a
number of kinematical cuts in Ref.\ \cite{nos} that we briefly present:

\begin{itemize}
  
\item [(C1)] We require that the leading jets and $e^\pm$ are in the
  pseudorapidity interval $|\eta| < 3$;
        
\item [(C2)] The leading leptons ($e_1$ and $e_2$) should have $p_T > 200$
  GeV;
      
\item [(C3)] We reject events where the invariant mass of the pair $e^+e^-$
  ($M_{e_1 e_2}$) is smaller than 190 GeV. This cut reduces the backgrounds
  coming from $Z$ decays into a pair $e^+ e^-$;
      
\item [(C4)] In order to further reduce the $t\bar{t}$ and remaining off-shell
  $Z$ backgrounds, we required that {\em all} the invariant masses $M_{e_i
    j_k}$ are larger than 200 GeV, since pairs $e_i j_k$ coming from an
  on-shell top decay have invariant masses smaller than $m_{\text top}$. The
  present experiments are able to search for leptoquarks with masses smaller
  than 200 GeV, therefore, this cut does not introduce any bias on the
  leptoquark search.

\end{itemize}

The above cuts reduce to a negligible level all the SM backgrounds \cite{nos}.
In principle we could also require the $e^\pm$ to be isolated from hadronic
activity in order to reduce the QCD backgrounds.  Nevertheless, we verified
that our results do not change when we introduce typical isolation cuts in
addition to any of the above cuts.  Since the leptoquark searches at the LHC
are free of backgrounds after these cuts \cite{nos}, the LHC will be able to
exclude with 95\% C.L.  the regions of parameter space where the number of
expected signal events is larger than 3 for a given integrated luminosity.

%%%%%%%%%%%%%%%%%%%%%%%%%%%%%%%%%%%%%%%%%%%%%%%%%%%%%%%%%%%%%%%%%%%%%%
\section{Results and Conclusions}
\label{resu}

\begin{figure}
\begin{center}
\mbox{\epsfig{file=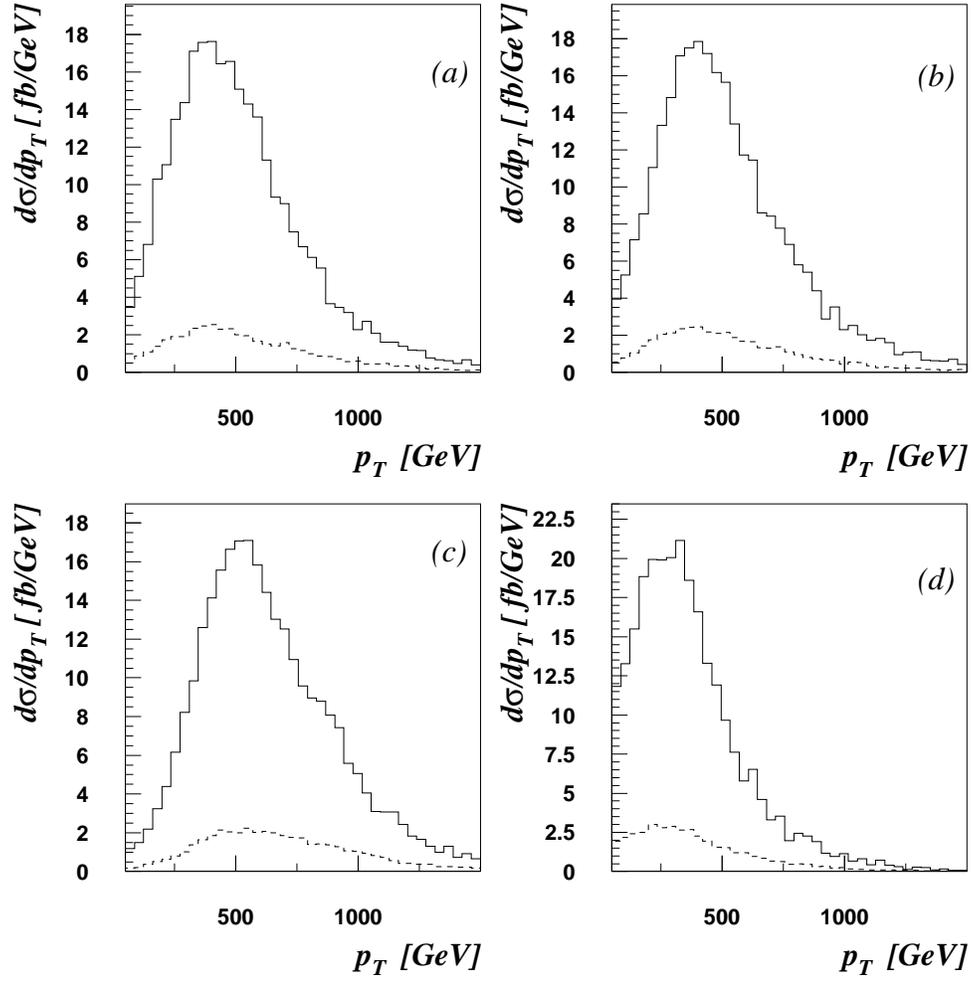,width=.85\linewidth}}
\end{center}
\caption{ $p_T$ distribution of (a) $e_1$; (b) $e_2$; (c) $j_1$; (d) $j_2$; 
  in the pair production of 1 TeV vector leptoquarks with $\beta=1$.  The
  dashed (continuous) line stands for the minimum (Yang--Mills) coupling.}
\label{fig1}
\end{figure}

\begin{figure}
\begin{center}
\mbox{\epsfig{file=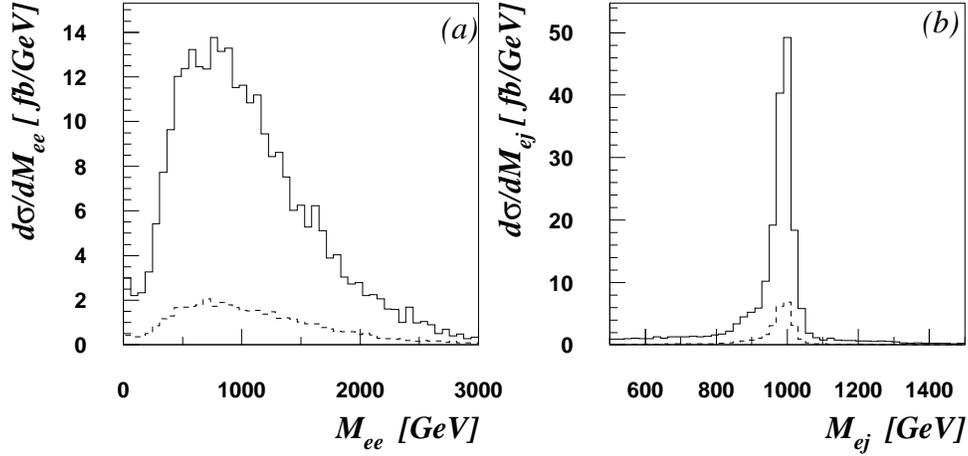,width=.85\linewidth}}
\end{center}
\caption{ (a) $e^+e^-$ invariant mass distribution; (b) $e^\pm$-jet 
  invariant mass spectrum adding the 4 possible combinations. We use the same
  conventions of Fig.\ \protect\ref{fig1}.}
\label{fig2}
\end{figure}

In order to access the effect of the cuts C1--C4 we exhibit in Fig.\ 
\ref{fig1} the $p_T$ distribution of the two most energetic leptons and jets
originating from the decay of a vector leptoquark of 1 TeV for minimal cross
section and Yang--Mills couplings to gluons. As we can see from this figure,
the $p_T$ distribution are peaked at $M_\Phi/2$ (=~500 GeV), and also exhibit
a large fraction of very hard jets and leptons. The presence of this peak
indicates that the two hardest jets and leptons usually originate from the
decay of the leptoquark pair. However, we still have to determine which are
the lepton and jet coming from the decay of one of the leptoquarks. Moreover,
we exhibit in Fig.\ \ref{fig2}a the $e^+ e^-$ invariant mass distribution
associated to 1 TeV vector leptoquark events. Clearly the bulk of the $e^+e^-$
pairs are produced at high invariant masses, and consequently the impact of
the cut C3 on the signal is small. Fig.\ \ref{fig2}b shows the invariant mass
distribution for the four possible $e_i j_k$ pairs combined in the 1 TeV
vector leptoquark case; the cut C4 does not affect significantly the signal
either.

\begin{figure}
\begin{center}
\mbox{\epsfig{file=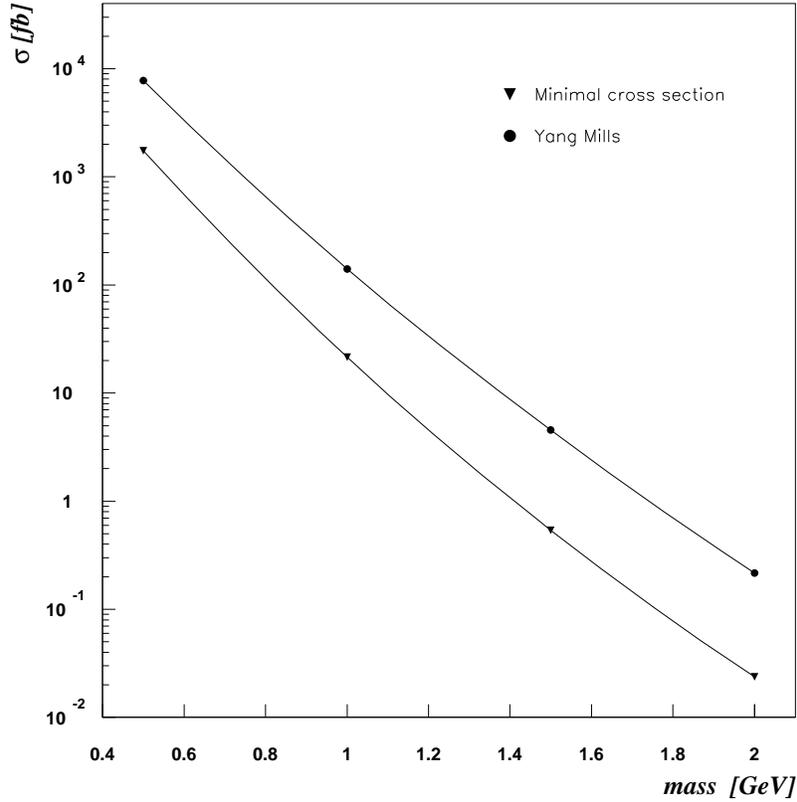,width=.7\linewidth}}
\end{center}
\caption{ Cross section after cuts for 
  the production of vector leptoquark pairs assuming Yang--Mills couplings
  (circles) and minimal cross section couplings (triangles).}
\label{fig:xsec}
\end{figure}

In our analyses of vector leptoquark pair production we applied the cuts
C1---C4 and also required the events to have two $e^\pm$-jet pairs with
invariant masses in the range $| M_\Phi \pm \Delta M|$ with $\Delta M$ given
in Table \ref{bins}. The pair production cross section after cuts is shown in
Fig.\ \ref{fig:xsec} for minimal cross section and Yang--Mills couplings.
For fixed values of $M_\Phi$, $\kappa_g$, and $\lambda_g$, the attainable
bounds at the LHC on vector leptoquarks depend upon its branching ratio
($\beta$) into a charged lepton and a jet, which is 0.5 or 1 for the
leptoquarks listed on Table \ref{t:cor}.

We exhibit in Table \ref{t:lim:pair} the 95\% C.L. limits on the leptoquark
masses that can be obtained from their pair production at the LHC for two
different integrated luminosities. In the worse scenario, {\em i.e.}  minimal
cross section couplings, the LHC will be able to place absolute bounds on
vector leptoquark masses smaller than 1.5 (1.6) TeV for $\beta = 0.5$ (1) and
an integrated luminosity of 10 fb$^{-1}$. With a larger luminosity of 100
fb$^{-1}$ this bound increases to 1.8 (1.9) TeV. Moreover, the limits are 300
GeV more stringent in the case of Yang--Mills coupling to gluons. At this
point it is interesting to compare our results with the ones in Ref.\ 
\cite{dion}. Requiring a $5\sigma$ signal as well as a minimum of 5 events
like in Ref.\ \cite{dion}, we obtain that the LHC will be able to rule out
vector leptoquarks with masses smaller than 2.0 (2.1) TeV for $\beta = 0.5$
(1), Yang--Mills couplings, and an integrated luminosity of 100 fb$^{-1}$.
Therefore, our cuts are more efficient than the ones proposed in Ref.\ 
\cite{dion} which lead to a bound of 1.55 TeV in the above conditions.

In brief, the discovery of vector leptoquarks is without any doubt a striking
signal of new physics beyond the standard model. The LHC will be of tremendous
help in the quest for new physics since, as we have shown, it will be able to
discover vector leptoquarks with masses smaller than 1.8--2.3 TeV, depending
in their couplings to fermions and gluons, through their pair production for
an integrated luminosity of 100 fb$^{-1}$.

%%%%%%%%%%%%%%%%%%%%%%%%%%%%%%%%%%%%%%%%%%%%%%%%%%%%%%%%%%%%%%%%%%%%%%
\acknowledgments

This work was partially supported by Conselho Nacional de Desenvolvimento
Cient\'\i fico e Tecnol\'ogico (CNPq), Funda\c{c}\~ao de Amparo \`a Pesquisa
do Estado de S\~ao Paulo (FAPESP), and by Programa de Apoio a N\'ucleos de
Excel\^encia (PRONEX).

%%%%%%%%%%%%%%%%%%%%%%%%%%%%%%%%%%%%%%%%%%%%%%%%%%%%%%

%%%tables

\begin{table}[htbp]
\begin{center}
\begin{tabular}{|c|c|c|}
leptoquark & decay & branching ratio 
\\
\hline
$V^1_{2L(R)} $         & $e^- d$ &  100\%  \\
$V^2_{2R} $            & $e^- u$ &  100\%  \\
$\tilde{V}^1_{2L}$     & $e^- u$ &  100\%  \\
$V_{1L}$               & $e^+ d$ &   50\%  \\
$V_{1R}$               & $e^+ d$ &  100\%  \\
$\tilde{V}_{1R}$       & $e^+ u$ &  100\%  \\
$V^-_{3L}$             & $e^+ u$ &  100\%  \\
$V^0_{3L}$             & $e^+ d$ &   50\%  \\
\end{tabular}
\vskip 0.75cm
\caption{Vector leptoquarks that can be observed through their decays into
  a $e^\pm$ and a jet and the correspondent branching ratios into this
  channel.}
\label{t:cor}
\end{center}
\end{table}

\begin{table}[htbp]
\begin{center}
\begin{tabular} {|c|c|c|c|c|}
Mass (GeV) &  Yang--Mills    &\multicolumn{3}{c|} {minimal cross section}   \\
\cline{3-5}
&  $\sigma$ (pb)  & $\sigma$ (pb) & $\kappa$ & $\lambda$\\
\hline
 500&28.5       &6.0            & 1.02   & -0.0409 \\
 600& 9.3       &1.8            & 1.06   & -0.0554 \\
 700& 3.4       &0.64           & 1.09   & -0.0691 \\
 800& 1.4       &0.24           & 1.12   & -0.0832 \\
 900& 0.61      &0.099          & 1.14   & -0.0967 \\
1000& 0.28      &0.043          & 1.17   & -0.111  \\
1100& 0.13      &0.019          & 1.24   & -0.153  \\   
1200& 0.066     &0.0091         & 1.24   & -0.163  \\   
1300& 0.033     &0.0044         & 1.25   & -0.175  \\  
1400& 0.017     &0.0021         & 1.26   & -0.187  \\  
1500& 0.0092    &0.0011         & 1.28   & -0.197  \\   
1600& 0.0049    &0.00056        & 1.29   & -0.208  \\  
1700& 0.0027    &0.00029        & 1.30   & -0.218  \\  
1800& 0.0014    &0.00015        & 1.30   & -0.227  \\  
1900& 0.00083   &0.00008        & 1.31   & -0.238  \\  
2000& 0.00046   &0.00004        & 1.32   & -0.247  \\ 
2100& 0.00026   &0.00002        & 1.32   & -0.256
\end{tabular}
\vskip 0.75cm
\caption{Total cross section in pb for the pair production of  vector 
  leptoquarks. The above values of $\kappa_g$ and $\lambda_g$ lead to a
  minimum value of the total cross section for a given leptoquark mass.}
\label{t:pair1}
\end{center}
\end{table}

%%%%%%

\begin{table}[htbp]
\begin{center}
\begin{tabular} {|c|c|}
$M_\Phi$ (GeV) & $\Delta M$ (GeV)
\\
\hline
 500   &  50 \\
1000   & 150 \\
1500   & 200 \\
2000   & 250 \\
\end{tabular}
\vskip 0.75cm
\caption{Invariant mass bins used in our analyses as a function of
  the leptoquark mass.}
\label{bins}
\end{center}
\end{table}

%%%

\begin{table}[htbp]
\begin{center}
\begin{tabular} {|c|c|c|}
       & minimal cross section & Yang--Mills \\
\hline
$V_{1L}$ and $V^0_{3L}$      & 1.5~(1.8)~TeV & 1.8~(2.1)~TeV \\
All others                   & 1.6~(1.9)~TeV & 1.9~(2.3)~TeV \\
\end{tabular}
\vskip 0.75cm
\caption{
  95\% CL limits on the leptoquark masses that can be obtained
  from the search for leptoquark pairs for two integrated
  luminosities ${\cal L}=$ 10 (100)~fb$^{-1}$ .  }

\label{t:lim:pair}
\end{center}
\end{table}

%%%%%%%%%Figures

%%%%%

\end{document}